\documentclass[runningheads]{llncs}
\usepackage{graphicx}
\usepackage{amsfonts}
\usepackage{amsmath}
\usepackage{subcaption}
\usepackage{bm}
\usepackage{url}
\usepackage{times}
\usepackage{multirow}
\usepackage{floatrow}
\newfloatcommand{capbtabbox}{table}[][\FBwidth]

\begin{document}
\title{Adaptive Honeypot Engagement through Reinforcement Learning of Semi-Markov Decision Processes}
\titlerunning{Adaptive Honeypot Engagement}
%
\author{Linan Huang \and
Quanyan Zhu\thanks{This research is supported in part by NSF under grant ECCS-1847056, CNS-1544782, and SES-1541164, and in part by ARO grant W911NF1910041.}}

%
\authorrunning{L. Huang and Q. Zhu}
%
\institute{Department of Electrical and Computer Engineering, New York University\\
 2 MetroTech Center, Brooklyn, NY, 11201, USA \\
\email{\{lh2328,qz494\}@nyu.edu}
}
\maketitle              
\begin{abstract}
A honeynet is a promising active cyber defense mechanism. It reveals the fundamental Indicators of Compromise (IoCs) by luring attackers to conduct adversarial behaviors in a controlled and monitored environment. 
The active interaction at the honeynet brings a high reward but also introduces high implementation costs and risks of adversarial honeynet exploitation.  
In this work, we apply infinite-horizon Semi-Markov Decision Process (SMDP) to characterize a stochastic transition and sojourn time of attackers in the honeynet and quantify the reward-risk trade-off. 
In particular, we design adaptive long-term engagement policies shown to be risk-averse, cost-effective, and time-efficient. 
Numerical results have demonstrated that our adaptive engagement policies can quickly attract  attackers to the target honeypot and engage them for a sufficiently long period to obtain worthy threat information. Meanwhile, the penetration probability is kept at a low level. 
 The results show that the expected utility is robust against attackers of a large range of persistence and intelligence. 
Finally, we apply reinforcement learning to the SMDP to solve the \textit{curse of modeling}. 
Under a prudent choice of the learning rate and exploration policy, we achieve a quick and robust convergence of the optimal policy and value. 

\end{abstract}
\section{Introduction}
Recent instances of \texttt{WannaCry} ransomware attack and \texttt{Stuxnet} malware have demonstrated an inadequacy of traditional cybersecurity techniques such as the firewall and intrusion detection systems. 
These passive defense mechanisms can detect low-level Indicators of Compromise (IoCs) 
such as hash values, IP addresses, and domain names. 
However, they can hardly disclose high-level indicators such as attack tools and Tactics, Techniques and Procedures (TTPs) of the attacker, which induces the attacker fewer pains to adapt to the defense mechanism, evade the indicators, and launch revised attacks as shown in the pyramid of pain \cite{WinNT}. 
Since high-level indicators are more effective in deterring emerging advanced attacks yet harder to acquire through the traditional passive mechanism, defenders need to adopt active defense paradigms to learn these fundamental characteristics of the attacker, attribute cyber attacks \cite{rid2015attributing}, and design defensive countermeasures correspondingly. 

Honeypots are one of the most frequently employed active defense techniques to gather information on threats. 
A honeynet is a network of honeypots, which emulates the real production system but has no production activities nor authorized services.  
Thus, an interaction with a honeynet, e.g., unauthorized inbound connections to any honeypot, directly reveals malicious activities.  
On the contrary, traditional passive techniques such as firewall logs or IDS sensors have to separate attacks from a ton of legitimate activities, thus provide much more false alarms and may still miss some unknown attacks. 

Besides a more effective identification and denial of adversarial exploitation through low-level indicators such as the inbound traffic, a honeynet can also help defenders to achieve the goal of identifying attackers' TTPs under proper engagement actions. The defender can interact with attackers and allow them to probe and perform in the honeynet until she has learned the attacker's fundamental characteristics. 
More services a honeynet emulates, more activities an attacker is allowed to perform, and a higher degree of interactions together result in a  larger revelation probability of the attacker's TTPs. 
However, the additional services and reduced restrictions also bring extra risks. Attacks may use some honeypots as  pivot nodes to launch attackers against other production systems \cite{spitzner2003honeypots}. 

\begin{figure}[]
\centering
\includegraphics[width=0.9 \textwidth]{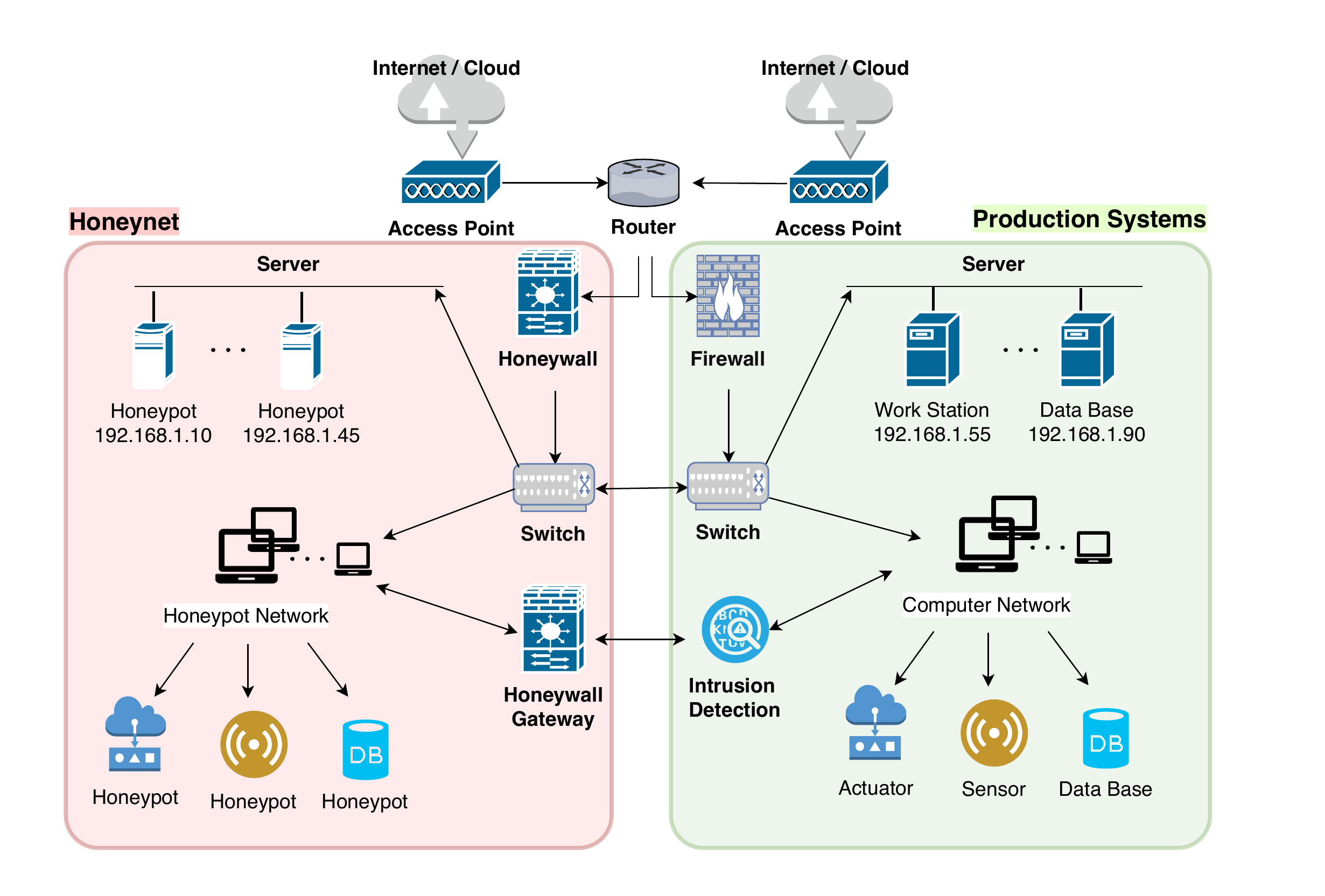}
\caption{
The honeynet in red mimics the targeted production system in green. The honeynet shares the same structure as the production system yet has no authorized services. 
 \label{fig: SystemStructure}}
\end{figure}
The current honeynet applies the honeywall as a gateway device to supervise outbound data and separate the honeynet from other production systems, as shown in Fig. \ref{fig: SystemStructure}. 
However, to avoid attackers' identification of the data control and the honeynet, a defender cannot block all outbound traffics from the honeynet, which leads to a trade-off between the rewards of learning high-level IoCs and the following three types of risks. 
 
\begin{itemize}
\item[T1:] Attackers identify the honeynet and thus either terminate on their own or generate misleading interactions with honeypots. 
\item[T2:] Attackers circumvent the honeywall to penetrate other production systems \cite{pouget2003white}. 
\item[T3:] Defender's engagement costs outweigh the investigation reward.  
\end{itemize}

We quantify risk T1 in Section \ref{sec:transition}, T2  in Section \ref{sec:riskT2}, and T3 in Section \ref{sec:investigationreward}. 
In particular, risk T3 brings the problem of timeliness and optimal decisions on timing. 
Since a persistent traffic generation to engage attackers is costly and the defender aims to obtain timely threat information, the defender needs cost-effective policies to lure the attacker quickly to the target honeypot and reduce attacker's sojourn time in honeypots of low-investigation value. 

\begin{figure}[]
\centering
\vspace{-10mm}
\includegraphics[width=0.85 \textwidth]{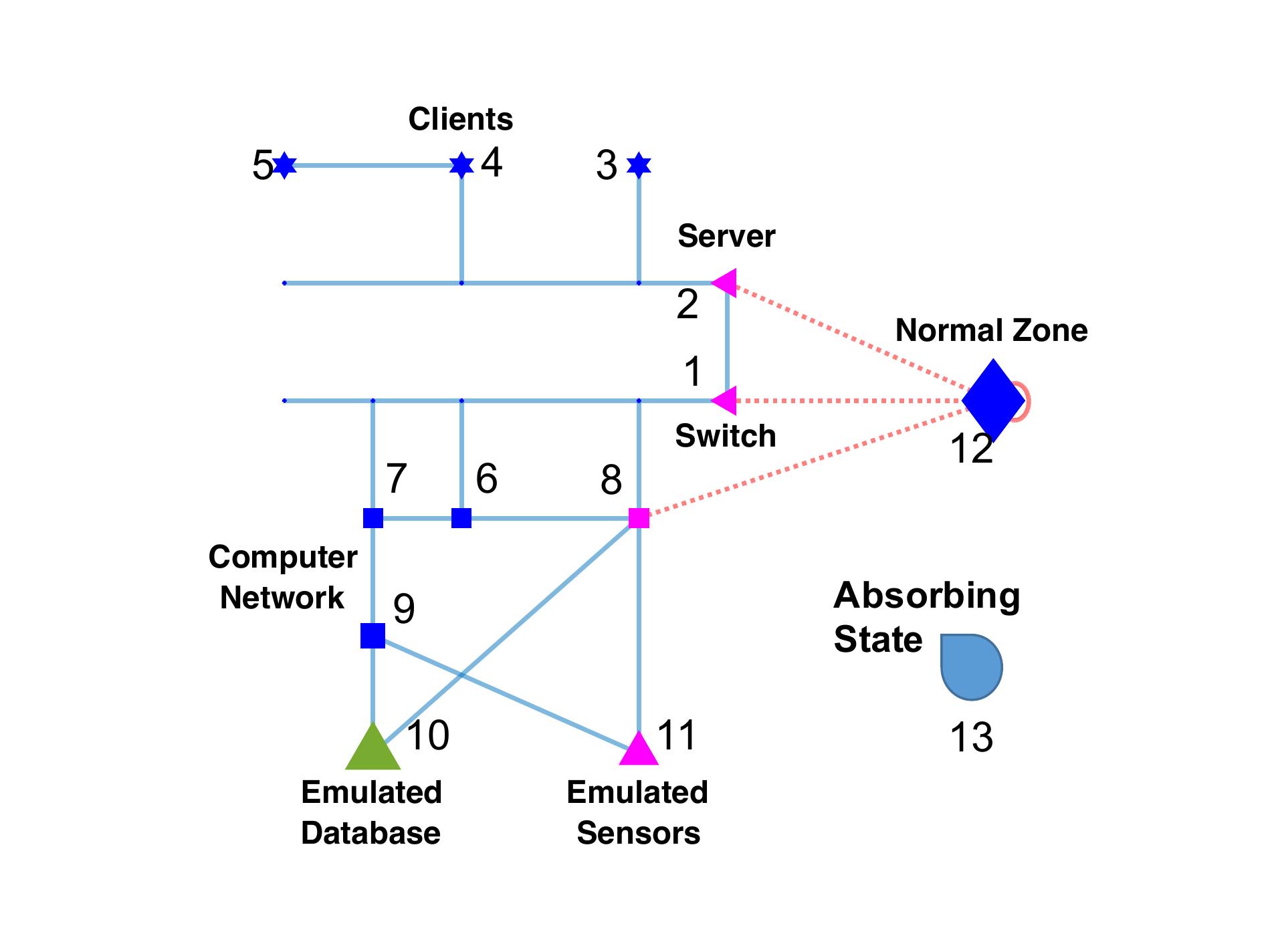}
\caption{
Honeypots emulate different components of the production system.  
}
 \label{fig: SMDPstructure}
\end{figure}

To achieve the goal of long-term, cost-effective policies,  we construct the Semi-Markov Decision Process (SMDP) in Section \ref{sec:formulation} on the network shown in Fig. \ref{fig: SMDPstructure}. 
Nodes $1$ to $11$ represent different types of honeypots, nodes $12$ and $13$ represent the domain of the production system and the virtual absorbing state, respectively. 
The attacker transits between these nodes according to the network topology in Fig. \ref{fig: SystemStructure} and can remain at different nodes for an arbitrary period of time. 
The defender can dynamically change the honeypots' engagement levels such as the amount of outbound traffic, to affect the attacker's sojourn time, engagement rewards, and the probabilistic transition in that honeypot. 

In Section \ref{sec:riskassessment}, we define security metrics related to our attacker engagement problem and analyze the risk both theoretically and numerically.  
These metrics answer important security questions in the honeypot engagement problem as follows.   
How likely will the attacker visit the normal zone at a given time? How long can a defender engage the attacker in a given honeypot before his first visit to the normal zone? How attractive is the honeynet if the attacker is initially in the normal zone?   
To protect against the Advanced Persistent Threats (APTs), we further investigate the engagement performance against attacks of different levels of persistence and intelligence. 

Finally, for systems with a large number of governing random variables, it is often hard to characterize the exact attack model, which is referred to as the \textit{curse of modeling}. 
Hence, we apply reinforcement learning methods in Section \ref{sec:reinforcementlearning} to learn the attacker's behaviors represented by the parameters of the SMDP. We visualize the convergence of the optimal engagement policy and the optimal value in a video demo\footnote{See the demo following URL: \url{https://bit.ly/2QUz3Ok}}. 
In Section \ref{sec:Discussion}, we discuss challenges and future works of reinforcement learning in the honeypot engagement scenario where the learning environment is non-cooperative, risky, and sample scarce.


\subsection{Related Works}
Active defenses \cite{mudrinich2012cyber} and defensive deceptions \cite{al2019autonomous} to detect and deter attacks have been active research areas. Techniques such as
honeynets \cite{jeffWiopt,zhu2013deployment}, moving target
defense \cite{zhu2013game,jajodia2011moving}, obfuscation \cite{pawlick_stackelberg_2016,pawlick2017mean},
and perturbations \cite{zhang_dynamic_2017,zhang2018distributed} have been introduced as defensive mechanisms to secure the cyberspace. 
The authors in \cite{horak2017manipulating} and \cite{huang2019dynamic} design two proactive defense schemes where the defender can manipulate the adversary's belief and take deceptive precautions under stealthy attacks, respectively.  
In particular, many works \cite{Hecker2012,pauna2018qrassh} including ones with Markov Decision Process (MDP) models \cite{luo2017iotcandyjar,jeffWiopt} and game-theoretic models \cite{wagener2009self,la2016deceptive,wang2017strategic} 
 focus on the adaptive honeypot deployment, configuration, and detection evasion to effectively gather threat  information without the attacker's notice.  A number of quantitative frameworks have been proposed to model proactive defense for various attack-defense scenarios building on Stackelberg games \cite{pawlick_stackelberg_2016,paruchuri2008playing,zhu2015game}, signaling games \cite{pawlick2017proactive,pawlick2018modeling,zhuang2010modeling,pawlick2015flip,xu2015cyber}, dynamic games \cite{sahabandu2018dift,huang2018PER,farhang2014dynamic,zhu2009dynamic}, and mechanism design theory \cite{chen2017security,zhang2017bi,hayel2015attack,zhu2012guidex}. Pawlick et al. in \cite{pawlick2019gameSURVEY} have provided a recent survey of game-theoretic methods for defensive deception, which includes a taxonomy of deception mechanisms and an extensive literature of game-theoretic deception.

Most previous works on honeypots have focused on studying the attacker's break-in attempts yet pay less attention to engaging the attacker after a successful penetration so that the attackers can thoroughly expose their post-compromise behaviors.  
Moreover, few works have investigated timing issues and risk assessment during the honeypot engagement, which may result in an improper engagement time and uncontrollable risks. 
The work most related to this one is \cite{jeffWiopt}, which introduces a continuous-state infinite-horizon MDP model where the defender decides when to eject the attacker from the network. The author assumes a maximum amount of information that a defender can learn from each attack. The type of systems, i.e., either a normal system or a honeypot,   determines the transition probability. 
Our framework, on the contrary, introduces following additional distinct features: 
\begin{itemize}
\item The upper bound on the amount of information which a defender can learn is hard to obtain and may not even exist. Thus, we consider a discounted factor to penalize the timeliness as well as the decreasing amount of unknown information as time elapses. 
\item The transition probability not only depends on the type of systems but also depends on the network topology and the defender's actions. 
\item The defender endows attackers the freedom to explore the honeynet and affects the transition probability and the duration time through different engagement actions. 
\item 
We use reinforcement learning methods to learn the parameter of the SMDP model. 
Since our learning algorithm constantly updates the engagement policy based on the up-to-date samples obtained from the honeypot interactions, the acquired optimal policy adapts to the potential evolution of attackers' behaviors. 
\end{itemize}

SMDP generalizes MDP by considering the random sojourn time at each state, and is widely applied to machine maintenance \cite{chen2005optimization}, resource allocation \cite{liang2012smdp}, infrastructure protection \cite{huang2018distributed,huang2018factored,huang2018distributed}, and cybersecurity \cite{sun2017update}. 
This work aims to leverage the SMDP framework to determine the optimal attacker engagement policy and to quantify the trade-off between the value of the investigation and the risk. 


\subsection{Notations}
Throughout the paper, we use calligraphic letter $\mathcal{X}$ to define a set. The upper case letter $X$ denotes a random variable and the lower case $x$ represents its realization. The boldface $\mathbf{X}$ denotes a vector or matrix and $\mathbf{I}$ denotes an identity matrix of a  proper dimension. 
Notation $\Pr$ represents the probability measure and 
$\star$ represents the convolution. 
 The indicator function $\mathbf{1}_{\{x=y\}}$ equals one if $x = y$, and zero if $x\neq y$. 
The superscript $k$ represents decision epoch $k$ and the subscript $i$ is the index of a node or a state. 
The pronoun `she' refers to the defender, and `he' refers to the attacker. 

\section{Problem Formulation}
\label{sec:formulation}
To obtain optimal engagement decisions at each honeypot under the probabilistic transition and the continuous sojourn time, we introduce the continuous-time infinite-horizon discounted SMDPs, which can be summarized by the tuple 
$ \{t\in [0,\infty), \mathcal{S},
\mathcal{A}({s_j}), \break
tr(s_l| s_j, a_j),  
z(\cdot| s_j,a_j,s_l), r^{\gamma}( s_j,a_j,s_l), \gamma\in [0,\infty)\}$. 
We describe each element of the tuple in this section. 
\subsection{Network Topology}
We abstract the structure of the honeynet as a finite graph $\mathcal{G}=(\mathcal{N},\mathcal{E})$. 
The node set $\mathcal{N}:=\{n_1,n_2,\cdots,n_N\}\cup \{n_{N+1}\}$ contains $N$ nodes of hybrid honeypots. Take Fig. \ref{fig: SMDPstructure} as an example, 
a node can be either a virtual honeypot of an integrated database system or a physical honeypot of an individual computer. 
These nodes provide different types of functions and services, and are connected following the topology of the emulated production system. 
Since we focus on optimizing the value of investigation in the honeynet, we only distinguish between different types of honeypots in different shapes, yet use one extra node $n_{N+1}$ to represent the entire domain of the production system. 
The network topology $\mathcal{E}:=\{e_{jl}\}, j,l\in \mathcal{N}$, is the set of directed links connecting node $n_j$ with $n_l$, and represents all possible transition trajectories in the honeynet. 
The links can be either physical (if the connecting nodes are real facilities such as computers) or logical (if the nodes represent integrated systems). 
Attackers cannot break the topology restriction. 
Since an attacker may use some honeypots as pivots to reach a production system, and it is also possible for a defender to attract attackers from the normal zone to the honeynet through these bridge nodes, there exist links of both directions between honeypots and the normal zone. 

\subsection{States and State-Dependent Actions}
At time $t\in [0,\infty)$, an attacker's state belongs to a finite set $\mathcal{S}:=\{s_1, s_2, \cdots, 
s_N, \allowbreak
s_{N+1},s_{N+2}\}$ where $s_i, i\in \{1,\cdots,N+1\}$, represents the attacker's location at time $t$. 
Once attackers are ejected or terminate on their own, we use the extra absorbing state $s_{N+2}$ to represent the virtual location. 
The attacker's state reveals the adversary visit and exploitation of the emulated functions and services. 
Since the honeynet provides a controlled environment, we assume that the defender can monitor the state and     transitions persistently without uncertainties. 
The attacker can visit a node multiple times for different purposes. A stealthy attacker may visit the honeypot node of the database more than once and revise data progressively (in a small amount each time) to evade detection. 
An attack on the honeypot node of sensors may need to frequently check the node for the up-to-date data. 
Some advanced honeypots may also emulate anti-virus systems or other protection mechanisms such as setting up an authorization expiration time, then the attacker has to compromise the nodes repeatedly.

At each state $s_i\in \mathcal{S}$,  the defender can choose an action $a_i$ from a state-dependent finite set $\mathcal{A}(s_i)$. 
For example, at each honeypot node, the defender can conduct action $a_E$ to eject the attacker, action $a_P$ to purely record the attacker's activities, low-interactive action  $a_L$, or high-interactive action $a_H$ to engage the attacker, i.e., $\mathcal{A}(s_i):=\{a_E, a_P,a_L,a_H\}, i\in \{1,\cdots,
\allowbreak
N\}$. 
The high-interactive action is costly to implement yet both increases the probability of a longer sojourn time at honeypot $n_i$, and reduces the probability of attackers penetrating the normal system from $n_i$ if connected. 
If the attacker resides in the normal zone either from the beginning or later through the pivot honeypots, the defender can choose either action $a_E$ to eject the attacker immediately, or action $a_A$ to attract the attacker to the honeynet by exposing some  vulnerabilities intentionally, i.e., $\mathcal{A}(s_{N+1}):=\{a_E,a_A\}$.  
Note that the instantiation of the action set and the corresponding consequences are not limited to the above scenario. For example, the action can also refer to a different degree of outbound data control. A strict control reduces the probability of attackers penetrating the normal system from the honeypot, yet also brings less investigation value. 
\subsection{Continuous-Time Process and Discrete Decision Model}
\label{sec:transition}
Based on the current state $s_j\in \mathcal{S}$, the defender's action $a_j\in \mathcal{A}(s_j)$, the attacker transits to state $s_l\in \mathcal{S}$ with a probability $tr(s_l|s_j,a_j)$ and the sojourn time at state $s_j$ is a continuous random variable with a probability density $z(\cdot| s_j,a_j, s_l)$. 
Note that the risk T1 of the attacker identifying the honeynet at state $s_j$ under action $a_j\neq A_E$
can be characterized by the transition probability $tr(s_{N+2}|s_j,a_j)$ as well as the duration time $z(\cdot| s_j,a_j,s_{N+2})$.  
Once the attacker arrives at a new honeypot $n_i$, the defender dynamically applies an interaction action at  honeypot $n_i$ from $\mathcal{A}(s_i)$ and keeps interacting with the attacker until he transits to the next honeypot. 
The defender may not change the action before the transition to reduce the probability of attackers detecting the change and  become aware of the honeypot engagement. 
Since the decision is made at the time of transition, we can transform the above continuous time model on horizon $t\in [0,\infty)$ into a discrete decision model at decision epoch $k\in \{0,1,\cdots, \infty\}$. 
The time of the attacker's $k^{th} $ transition is denoted by a random variable $T^k$, the landing state is denoted as $s^k\in \mathcal{S}$, and the adopted action  after arriving at $s^k$ is denoted as $a^k\in \mathcal{A}(s^k)$. 

\subsection{Investigation Value}
\label{sec:investigationreward}
The defender gains a reward of investigation by engaging and analyzing the attacker in the honeypot. 
To simplify the notation, we divide the reward during time $t\in [0,\infty)$ into ones at discrete decision epochs $T^k, k\in \{0,1,\cdots, \infty\}$. 
When $\tau\in [T^k,T^{k+1}]$ amount of time elapses at stage $k$, the defender's reward of investigation
\begin{equation*}
r(s^k,a^k,s^{k+1},T^k,T^{k+1},\tau)=r_1(s^k,a^k,s^{k+1})\mathbf{1}_{\{\tau=0\}}+r_2(s^k,a^k, T^k,T^{k+1},\tau), 
\end{equation*}
at time $\tau$ of stage $k$, is the sum of two parts. The first part is the immediate cost of applying engagement action $a^k \in \mathcal{A}(s^k)$ at state $s^k \in \mathcal{S}$ and the second part is the reward rate of threat information acquisition minus the cost rate of persistently generating deceptive traffics. 
Due to the randomness of the attacker's behavior, the information acquisition can also be random, thus the actual reward rate $r_2$ is perturbed by an additive zero-mean noise $w_r$. 

Different types of attackers target different components of the production system. For example, an attacker who aims to steal data will take intensive adversarial actions at the database.
Thus, if the attacker is actually in the honeynet and adopts the same behavior as he is in the production system, the defender can identify the target of the attack based on the traffic intensity. 
We specify $r_1$ and $r_2$ at each state properly to measure the risk T3. 
To maximize the value of the investigation, the defender should choose proper actions to lure the attacker to the honeypot emulating the target of the attacker in a short time and with a large probability. 
Moreover, the defender's action should be able to engage the attacker in the target honeypot actively for a longer time to obtain more valuable threat information. 
We compute the optimal long-term policy that achieves the above objectives in Section \ref{sec:optimalpolicy}. 

As the defender spends longer time interacting with attackers, investigating their behaviors and acquires better understandings of their targets and TTPs, less new information can be extracted. 
In addition, the same intelligence becomes less valuable as time elapses due to the  timeliness. 
Thus, we use a discounted factor of $\gamma\in [0,\infty)$ to penalize the decreasing value of the investigation as time elapses. 

\subsection{Optimal Long-Term Policy}
\label{sec:optimalpolicy}
The defender aims at a policy $\pi\in \Pi$ which maps state $s^k\in \mathcal{S}$ to action $a^k\in \mathcal{A}(s^k)$ to maximize the long-term expected utility starting from state $s^0$, i.e., 
\begin{align*}
u(s^0,\pi)=\mathbb{E}\left[\sum_{k=0}^{\infty} \int_{T^k}^{T^{k+1}} e^{-\gamma(\tau+T^k)} (r(S^k,A^k,S^{k+1},T^k,T^{k+1},\tau)+w_r)d\tau\right]. 
\end{align*}

At each decision epoch, the value function $v(s^0)=\sup_{\pi \in \Pi }u(s^0,\pi)$ can be represented by dynamic programming, i.e., 
\begin{align}
\label{eq:DPgeneral}
v(s^0)=\sup_{a^0\in \mathcal{A}(s^0)} \mathbb{E}\left[\int_{T^0}^{T^{1}} e^{-\gamma (\tau+T^0)}r(s^0,a^0,S^{1},T^0,T^{1},\tau)d\tau+e^{-\gamma T^1}v(S^1)\right]. 
\end{align}

We assume a constant reward rate $r_2(s^k,a^k,T^k,T^{k+1},\tau)=\bar{r}_2(s^k,a^k)$ for simplicity. 
Then, \eqref{eq:DPgeneral} can be transformed into an  equivalent  MDP form, i.e., $ \forall s^0\in \mathcal{S}$, 
\begin{align}
v(s^0)=\sup_{a^0\in \mathcal{A}(s^0)} \sum_{s^1\in \mathcal{S}} tr(s^1|s^0,a^0) (r^{\gamma}(s^0,a^0,s^1)+z^{\gamma}(s^0,a^0,s^1)v(s^1)), 
\end{align}
where ${z^{\gamma}}(s^0,a^0,s^1):=\int_0^{\infty}e^{-\gamma \tau} z(\tau|s^0,a^0,s^1)d\tau\in [0,1]$ is the Laplace transform of the sojourn probability density $z(\tau|s^0,a^0,s^1)$ and the equivalent reward 
$r^{\gamma}(s^0,a^0,s^1)\allowbreak
:=r_1(s^0,a^0,s^1)+\frac{\bar{r}_2(s^0,a^0)}{\gamma} (1-z^{\gamma}(s^0,a^0,s^1))\in [-m_c,m_c]$ is assumed to be bounded by a constant $m_c$. 

A classical regulation condition of SMDP to avoid the probability of an infinite number of transitions within a finite time is stated as follows:  there exists constants $\theta\in (0,1)$ and $\delta>0$ such that 
\begin{align}
\label{eq:ReguCondition}
\sum_{s^1\in \mathcal{S}} tr(s^1|s^0,a^0) z(\delta |s^0,a^0,s^1)\leq 1-\theta , \forall s^0\in \mathcal{S}, a^0 \in \mathcal{A}(s^0). 
\end{align}
It is shown in \cite{hu2007markov} that condition \eqref{eq:ReguCondition} is equivalent to  
$
\sum_{s^1\in \mathcal{S}}  tr(s^1|s^0,a^0) z^{\gamma}(s^0,a^0,s^1)\in [0,1), 
$
which serves as the equivalent  stage-varying discounted factor for the associated MDP. 
Then, the right-hand side of \eqref{eq:DPgeneral} is a contraction mapping and there exists a unique optimal policy $\pi^*=arg\max_{\pi \in \Pi }u(s^0,\pi)$ which can be found by value iteration, policy iteration or linear programming. 

\subsubsection{Cost-Effective Policy}
The computation result of our $13$-state example system is illustrated in Fig. \ref{fig: SMDPstructure}. 
The optimal policies at honeypot nodes $n_1$ to $n_{11}$ are represented by different colors. Specifically, actions $a_E,a_P,a_L,a_H$ are denoted in red, blue, purple, and green, respectively. The size of node $n_i$ represents the state value $v(s_i)$.  

In the example scenario, the honeypot of database $n_{10}$ and  sensors  $n_{11}$ are the main and secondary targets of the attacker, respectively. Thus, defenders can obtain a higher investigation value when they manage to engage the attacker in these two honeypot nodes with a larger probability and for a longer time. 
However, instead of naively adopting high interactive actions,  
a savvy defender also balances the high implantation cost of $a_H$. Our quantitative results indicate that the high interactive action should only be applied at $n_{10}$ to be cost-effective. 
On the other hand, although the bridge nodes $n_1,n_2,n_8$ which connect to the normal zone $n_{12}$ do not contain higher investigation values than other nodes, the defender still takes action $a_L$  at these nodes. 
The goal is to either increase the probability of attracting attackers away from the normal zone or reduce the probability of attackers penetrating the normal zone from these bridge nodes. 

\subsubsection{Engagement Safety versus Investigation Values}
\label{sec:riskT2}
Restrictive engagement actions endow attackers less freedom so that they are less likely to penetrate the normal zone. However, restrictive actions also decrease the probability of obtaining high-level IoCs, thus reduces the investigation values. 

\begin{figure}[]
\RawFloats
\centering
\vspace{-8mm}
\includegraphics[width=0.9 \textwidth]{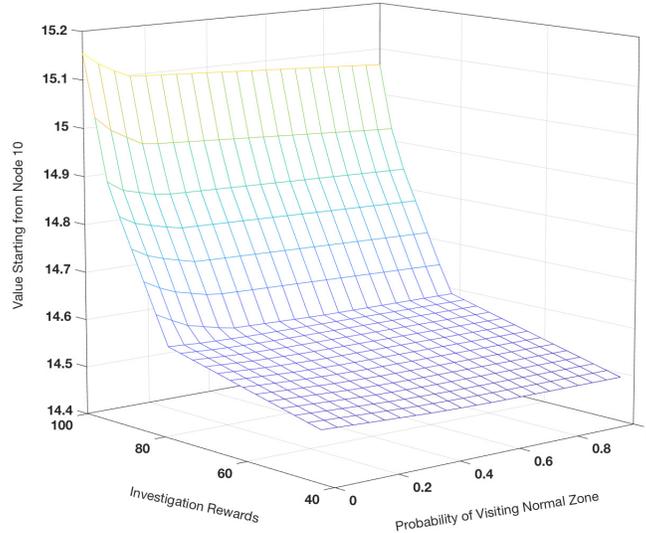}
\caption{
The trade-off surface of $v(s_{10})$ in $z$-axis under different values of penetration probability $p(s_{N+1}|s_j,a_j), j\in \{s_1,s_2,s_8\}, a_j \neq a_E$, in $x$-axis, and the reward $r^{\gamma}(s_j,a_j,s_l), j\in \mathcal{S}\setminus \{s_{12},s_{13}\},l \in \mathcal{S}$, in $y$-axis. 
 \label{fig: tradeoff}}
\end{figure}

To quantify the system value under the trade-off of the engagement safety and the reward from the investigation, we visualize the trade-off surface in Fig. \ref{fig: tradeoff}. 
In the $x$-axis, a larger penetration probability $p(s_{N+1}|s_j,a_j), j\in \{s_1,s_2,s_8\}, a_j \neq a_E$, decreases the value $v(s_{10})$. In the $y$-axis, a larger  reward $r^{\gamma}(s_j,a_j,s_l), j\in \mathcal{S}\setminus \{s_{12},s_{13}\},l \in \mathcal{S}$, increases the value. 
The figure also shows that value $v(s_{10})$ changes in a higher rate, i.e., are more sensitive when the penetration probability is small and the reward from the investigation is large. 
In our scenario, the penetration probability has less influence on the value than the investigation reward, which motivates a less restrictive engagement. 

\section{Risk Assessment}
\label{sec:riskassessment}
Given any feasible engagement policy $\pi\in \Pi$, the SMDP becomes a semi-Markov process \cite{nakagawa2011stochastic}. 
We analyze the evolution of the occupancy distribution and first passage time in Section \ref{sec:occupancy} and \ref{sec:firstpassagetime}, respectively, which leads to three security metrics during the honeypot engagement. 
To shed lights on the defense of APTs, we investigate the system performance against attackers with different levels of persistence and intelligence in Section \ref{sec:APT}.  
\subsection{Transition Probability of Semi-Markov Process} 
\label{sec:occupancy}
Define the cumulative probability $q_{ij}(t)$ of the one-step transition  from $\{S^k=i,T^k=t^k\}$ to $\{S^{k+1}=j,T^{k+1}=t^k+t\}$ as $\Pr(S^{k+1}=j,T^{k+1}-t^k\leq t| S^k=i,T^k=t^k)=tr(j|i,\pi(i))  \int_0^t z(\tau|i,\pi(i),j)d\tau, \forall i,j\in \mathcal{S},t\geq 0$. 
Based on a variation of the forward Kolmogorov equation where the one-step transition lands on an intermediate state $l\in \mathcal{S}$ at time $T^{k+1}=t^k+u, \forall u\in [0,t]$, the transition probability of the system in state $j$ at time $t$, given the initial state $i$ at time $0$ can be represented as 
\begin{align*}
& p_{ii}(t)=1-\sum_{h\in \mathcal{S}} q_{ih}(t) +\sum_{l\in \mathcal{S}} \int_0^t p_{li}(t-u)dq_{il}(u),  \\
& p_{ij}(t)=\sum_{l\in \mathcal{S}} \int_0^t p_{lj}(t-u)dq_{il}(u)=\sum_{l\in \mathcal{S}}  p_{lj}(t) \star \frac{dq_{il}(t)}{dt}, \forall i,j\in \mathcal{S}, j\neq i, \forall t\geq 0, 
\end{align*}
where $1-\sum_{h\in \mathcal{S}} q_{ih}(t)$  is the probability that no transitions happen before time $t$. 
We can easily verify that $\sum_{l\in \mathcal{S}}p_{il}(t)=1, \forall i\in \mathcal{S}, \forall t\in [0,\infty)$.
To compute $p_{ij}(t)$ and $p_{ii}(t)$, we can take Laplace transform and then solve two sets of linear equations. 

For simplicity, we specify $z(\tau| i,\pi(i),j)$ to be exponential distributions with parameters $\lambda_{ij}(\pi(i))$, and the semi-Markov process degenerates to a continuous time Markov chain. 
Then, we obtain the infinitesimal generator via the Leibniz integral rule, i.e., 
\begin{align*}
& \bar{q}_{ij}:=\frac{d p_{ij}(t)}{dt} \bigg|_{t=0}=\lambda_{ij}(\pi(i)) \cdot tr(j|i,\pi(i))>0, \forall i,j \in \mathcal{S},  j\neq i, 
\\
& \bar{q}_{ii}:=\frac{d p_{ii}(t)}{dt} \bigg|_{t=0}=-\sum_{j\in \mathcal{S}\setminus \{i\}} \bar{q}_{ij}<0, \forall i\in \mathcal{S}. 
\end{align*}
Define matrix $\mathbf{\bar{Q}}:=[\bar{q}_{ij}]_{i,j\in \mathcal{S}}$ and vector $\mathbf{P}_i(t)=[p_{ij}(t)]_{j\in \mathcal{S}}$, then based on the forward Kolmogorov equation,
\begin{align*}
\frac{d\mathbf{P}_i(t)}{dt}=\lim_{u\rightarrow 0^+}\frac{\mathbf{P}_i(t+u)-\mathbf{P}_i(t)}{u}=\lim_{u\rightarrow 0^+}\frac{\mathbf{P}_i(u)-\mathbf{I} }{u}\mathbf{P}_i(t)=\mathbf{\bar{Q}}\mathbf{P}_i(t). 
\end{align*}
Thus, we can compute the first security metric, the \textit{occupancy distribution} of any state $s\in \mathcal{S}$ at time $t$ starting from the initial state $i\in \mathcal{S}$ at time $0$, i.e., 
\begin{equation}
\label{eq:Q2P}
\mathbf{P}_i(t)=e^{\mathbf{\bar{Q}}t}\mathbf{P}_i(0), \forall i\in \mathcal{S}. 
\end{equation}

\begin{figure}[]
\RawFloats
\centering
\minipage[t]{0.48\textwidth}
\includegraphics[width=\textwidth]{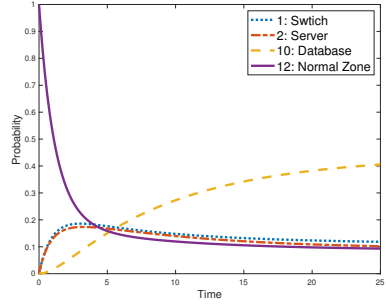}
\caption{
Evolution of $p_{ij}(t), i=s_{N+1}$. 
 \label{fig: DensityEvolve}}
\endminipage
\hfill
\minipage[t]{0.51\textwidth}
\centering
\includegraphics[width=1 \textwidth]{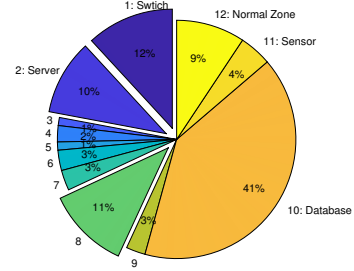}
\caption{
The limiting occupancy distribution. 
 \label{fig: piechart}}
\endminipage
\end{figure}

We plot the evolution of $p_{ij}(t), i=s_{N+1}, j\in \{s_1,s_2,s_{10},s_{12}\}$, versus $t\in [0,\infty)$ in Fig. \ref{fig: DensityEvolve} and  the limiting occupancy distribution $p_{ij}(\infty), i=s_{N+1}$, in Fig. \ref{fig: piechart}. 
In Fig. \ref{fig: DensityEvolve}, although the attacker starts at the normal zone $i=s_{N+1}$, our engagement policy can quickly attract the attacker into the honeynet. 
Fig. \ref{fig: piechart} demonstrates that the engagement policy can keep the attacker in the honeynet with a dominant probability of $91\%$ and specifically, in the target honeypot $n_{10}$ with a high probability of $41\%$. 
The honeypots connecting the normal zone also have a higher occupancy probability than nodes $n_3,n_4,n_5,n_6,n_7,n_9$, which are less likely to be explored by the attacker due to the network topology. 

%


\subsection{First Passage Time}
\label{sec:firstpassagetime}
Another quantitative measure of interest is the first passage time $T_{i\mathcal{D}}$ of visiting a set $\mathcal{D}\subset \mathcal{S}$ starting from $i\in \mathcal{S}\setminus \mathcal{D}$ at time $0$. 
Define the cumulative probability function $f^c_{i\mathcal{D}}(t):=\Pr(T_{i\mathcal{D}}\leq t)$, then 
$
f^c_{i\mathcal{D}}(t)=\sum_{h\in \mathcal{D}} q_{ih}(t)+\sum_{l \in \mathcal{S}\setminus \mathcal{D}}\int_0^t f^c_{l\mathcal{D}}(t-u)dq_{il}(u). 
$
In particular, if $\mathcal{D}=\{j\}$, then the probability density function $f_{ij}(t):=\frac{df^c_{ij}(t)}{dt}$ satisfies 
\begin{equation*}
p_{ij}(t)=\int_0^t p_{jj}(t-u)df^c_{ij}(u)=p_{jj}(t)\star f_{ij}(t), \forall i,j \in \mathcal{S}, j\neq i.
\end{equation*}
Take Laplace transform $\bar{p}_{ij}(s):=\int_0^{\infty}e^{-s t}p_{ij}(t)dt$, and then take inverse Laplace transform on $\bar{f}_{ij}(s)=\frac{\bar{p}_{ij}(s)}{\bar{p}_{jj}(s)}$, we obtain
\begin{align}
\label{eq:PDFofTD}
f_{ij}(t)=\int_0^{\infty}e^{s t}\frac{\bar{p}_{ij}(s)}{\bar{p}_{jj}(s)} ds,\forall i,j \in \mathcal{S}, j\neq i. 
\end{align}

\begin{figure}
\RawFloats
\centering
\includegraphics[width=0.55 \textwidth]{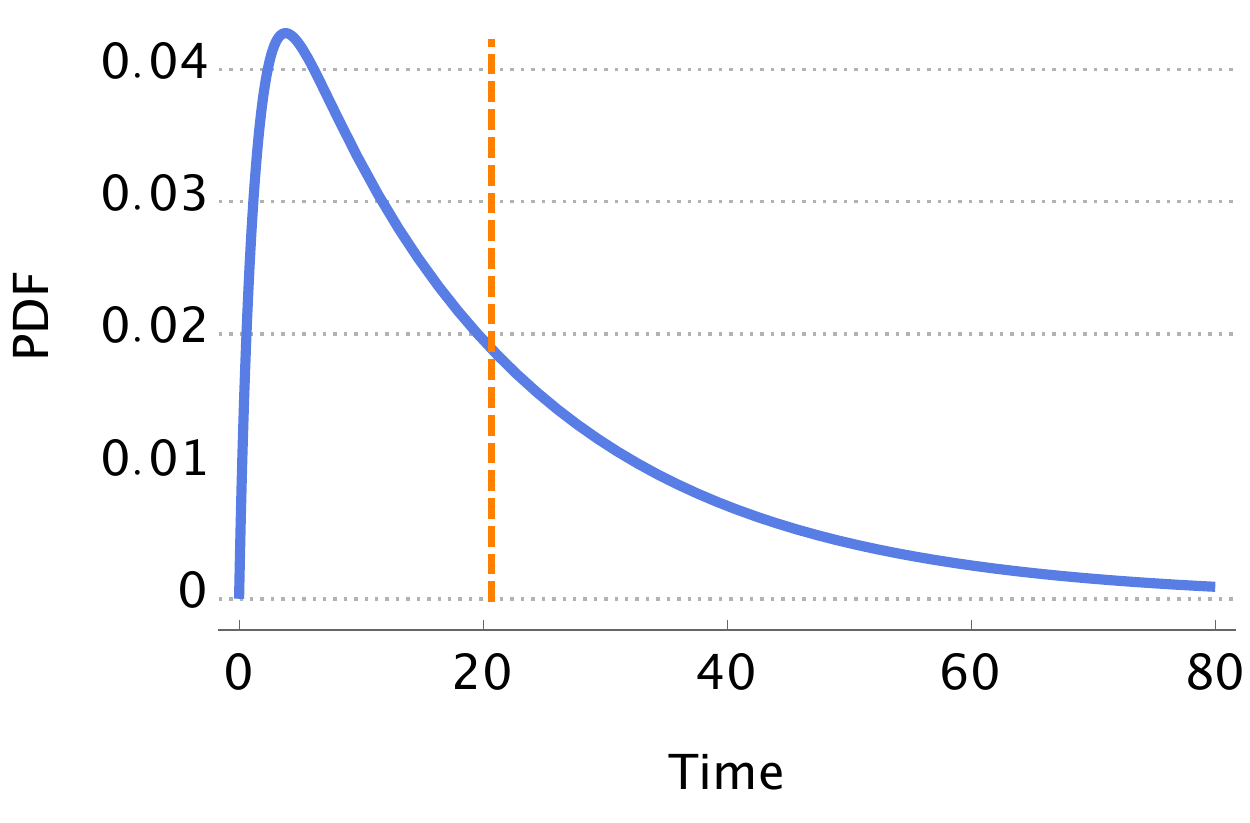}
\caption{
Probability density function of $T_{s_{12},s_{10}}$. 
 \label{fig: PDF}}
\end{figure}

We define the second security metric, the  \textit{attraction efficiency} as the probability of the first passenger time $T_{s_{12},s_{10}}$ less than a threshold $t_{th}$. 
Based on \eqref{eq:Q2P} and \eqref{eq:PDFofTD}, the probability density function of $T_{s_{12},s_{10}}$ is shown in Fig. \ref{fig: PDF}. We take the mean denoted by the orange line as the threshold $t_{th}$ and the attraction efficiency is $0.63 $, which means that the defender can attract the attacker from the normal zone to the database honeypot in less than $t_{th}=20.7$ with a probability of $0.63$.

\subsubsection{Mean First Passage Time}
The third security metric of concern is the \textit{average engagement efficiency} defined as the Mean First Passage Time (MFPT) $t^{m}_{i \mathcal{D}}=E[T_{i\mathcal{D}}],\forall i\in \mathcal{S}, \mathcal{D}\subset \mathcal{S}$. 
Under the exponential sojourn distribution, MFPT can be computed directly through the a system of linear equations, i.e.,  
\begin{equation}
\label{eq:MFPT}
\begin{split}
& t_{iD}^m=0, i\in \mathcal{D}, \\
& 1+\sum_{l\in \mathcal{S}} \bar{q}_{il}t^{m}_{l\mathcal{D}}=0, i\notin \mathcal{D}. 
\end{split}
\end{equation} 

\begin{figure}
\RawFloats
\centering
\minipage[t]{0.48\textwidth}
\includegraphics[width=\textwidth]{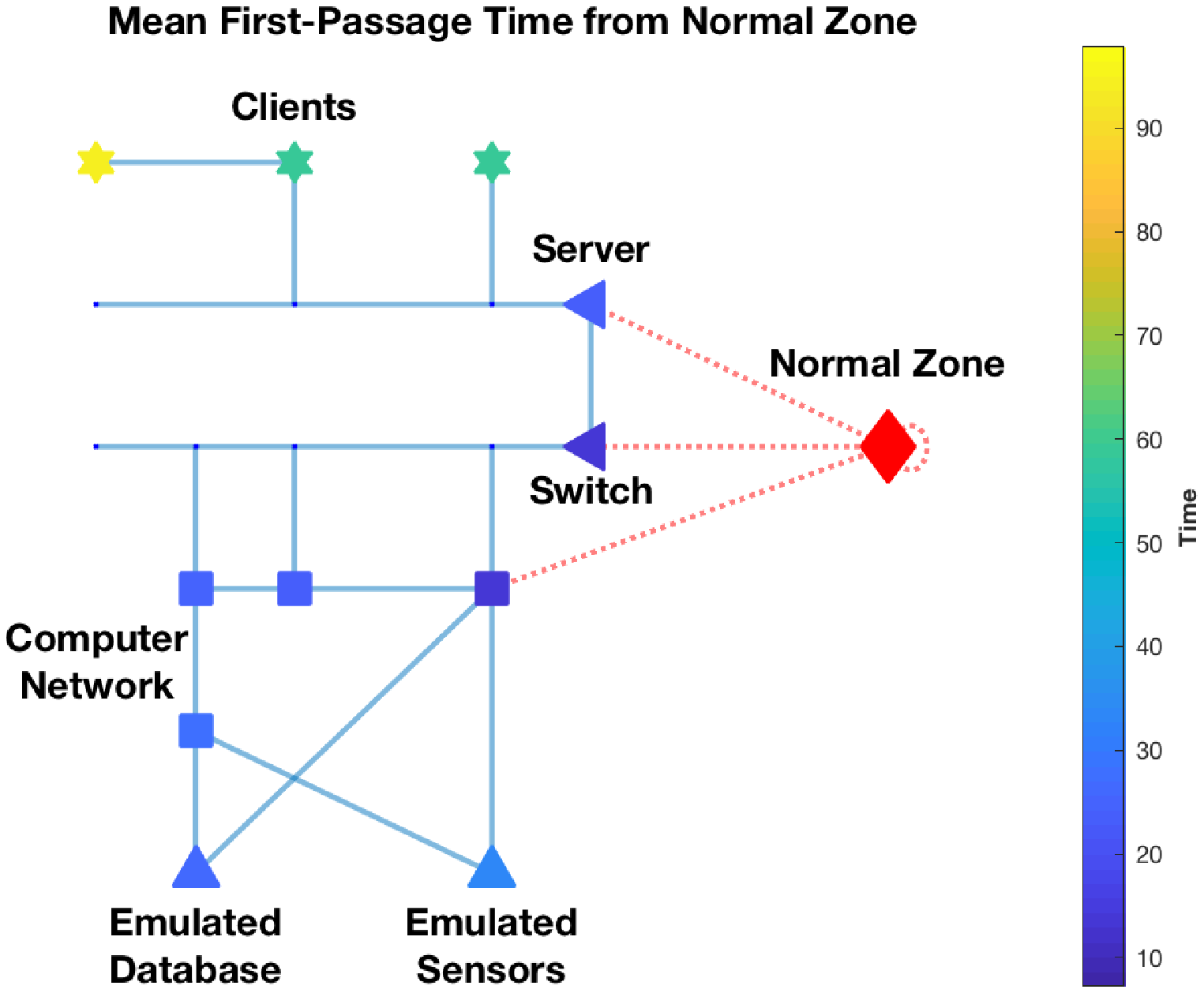}
\caption{
MFPT from the normal zone $t^m_{s_{12},j}$.
 \label{fig: MFPTfrom12}}
\endminipage
\hfill
\minipage[t]{0.48\textwidth}
\centering
\includegraphics[width= \textwidth]{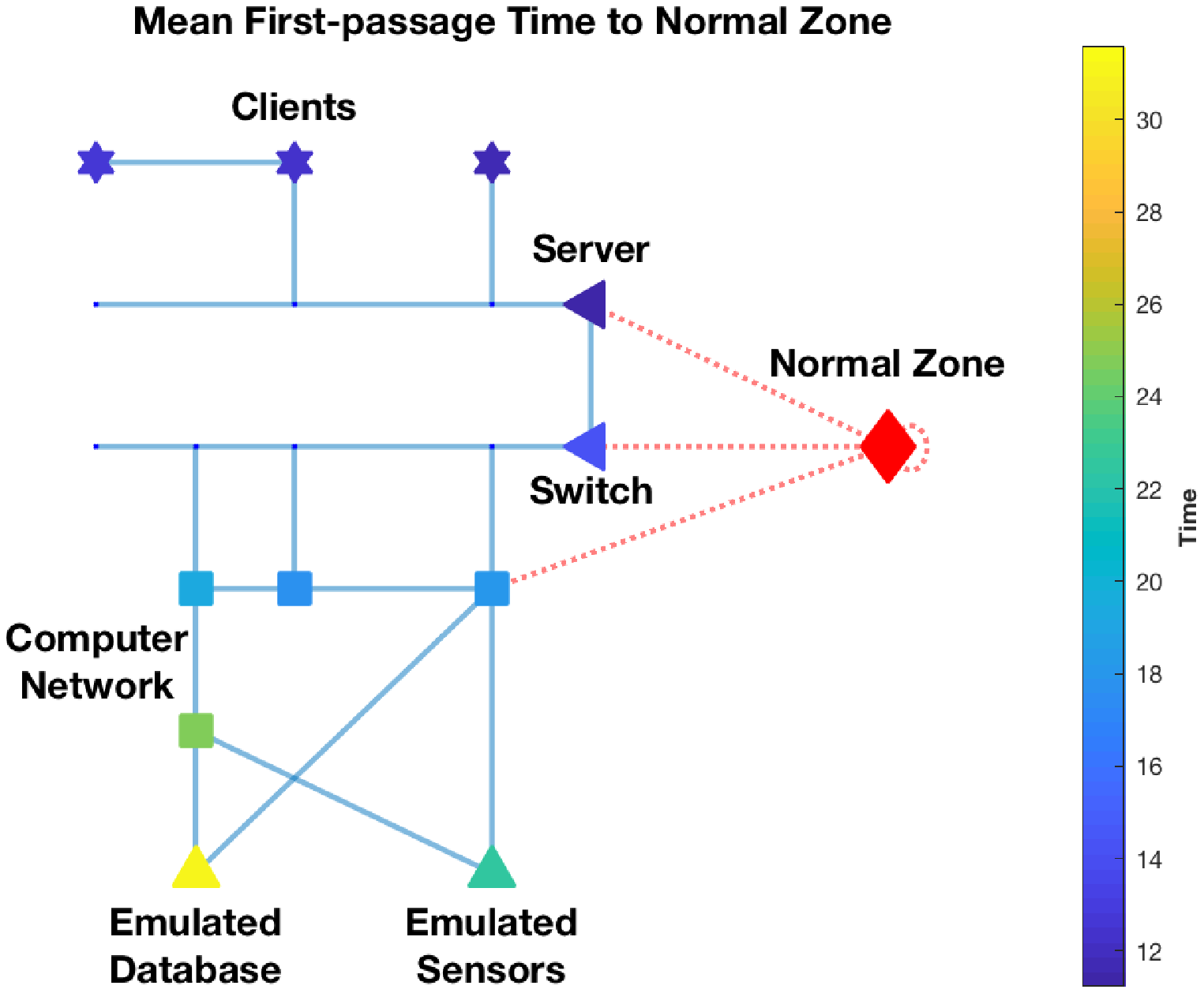}
\caption{
MFPT to the normal zone $t^m_{i, s_{12}}$.
 \label{fig: MFPTto12}}
\endminipage
\end{figure}

In general,  the MFPT is asymmetric, i.e., $t^m_{ij}\neq t^m_{ji}, \forall i,j\in \mathcal{S}$. 
Based on \eqref{eq:MFPT}, we compute the MFPT from and to the normal zone in Fig. \ref{fig: MFPTfrom12} and Fig. \ref{fig: MFPTto12}, respectively. The color of each node indicates the value of MFPT. 
In Fig. \ref{fig: MFPTfrom12}, the honeypot nodes that directly connect to the normal zone have the shortest MFPT, and it takes attackers much longer time to visit the honeypots of clients due to the network topology.  
Fig. \ref{fig: MFPTto12} shows that the defender can engage attackers in the target honeypot nodes of database and sensors for a longer time. The engagements at the client nodes are yet much less attractive. 
Note that two figures have different time scales denoted by the color bar value, and the comparison shows that it generally takes the defender more time and efforts to attract the attacker from the normal zone. 

The MFPT from the normal zone $t^m_{s_{12},j}$ measures the average time it takes to attract attacker to honeypot state $j\in \mathcal{S}\setminus \{s_{12},s_{13}\}$ for the first time. On the contrary, the MFPT to the normal zone  $t^m_{i,s_{12}}$ measures the average time of the attacker penetrating the normal zone from honeypot state $i\in \mathcal{S}\setminus \{s_{12},s_{13}\}$ for the first time. 
If the defender pursues absolute security and ejects the attack once it goes to the normal zone, then Fig. \ref{fig: MFPTto12} also shows the attacker's average sojourn time in the honeynet starting from different honeypot nodes. 

\subsection{Advanced Persistent Threats}
\label{sec:APT}
\begin{figure}
\RawFloats
\centering
\minipage[t]{0.47\textwidth}
\includegraphics[width=1\textwidth]{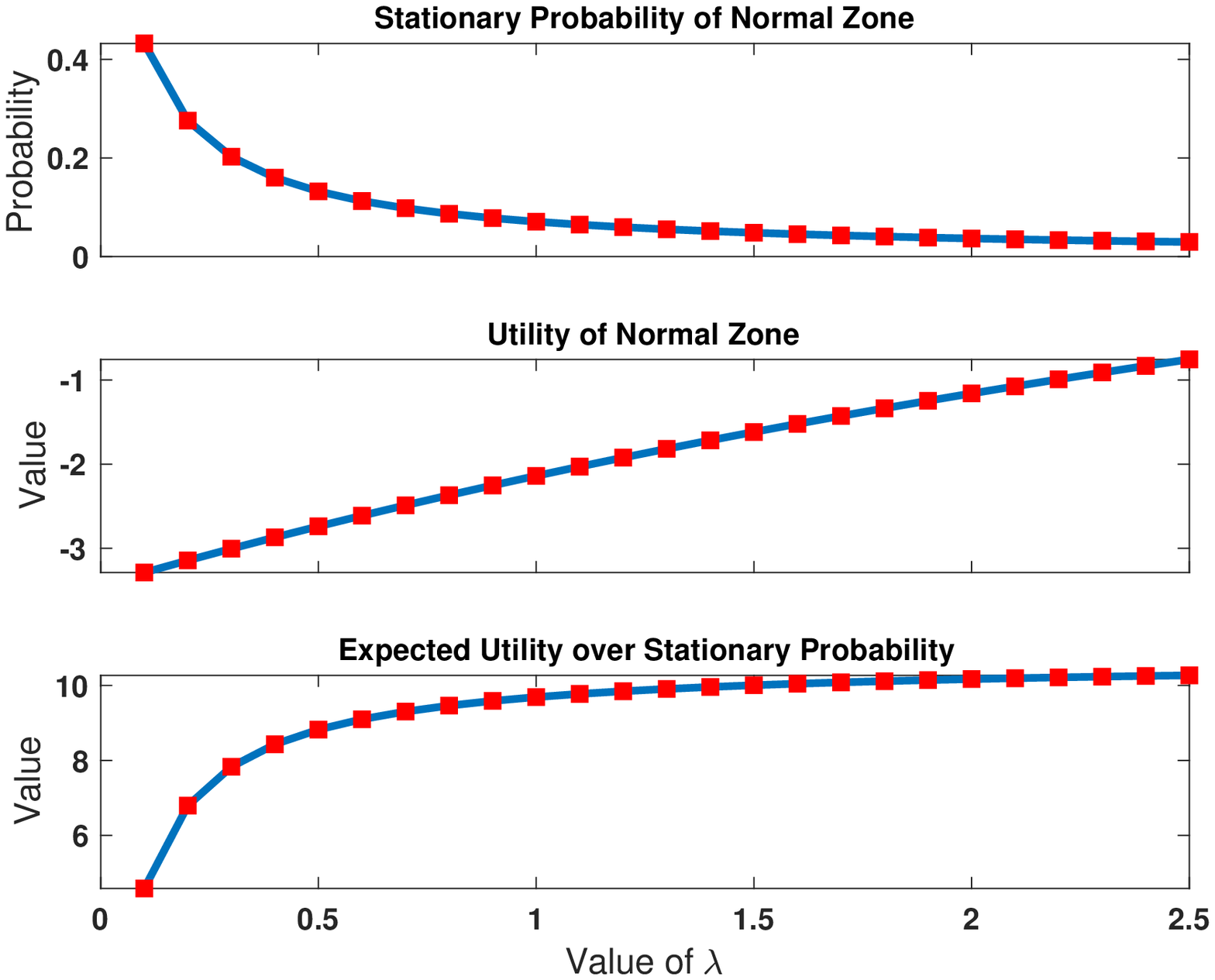}
\caption{
Three engagement criteria under different persistence levels $\lambda\in (0,2.5]$.
 \label{fig: persistency}}
\endminipage
\hfill
\minipage[t]{0.47\textwidth}
\centering
\includegraphics[width= 1\textwidth]{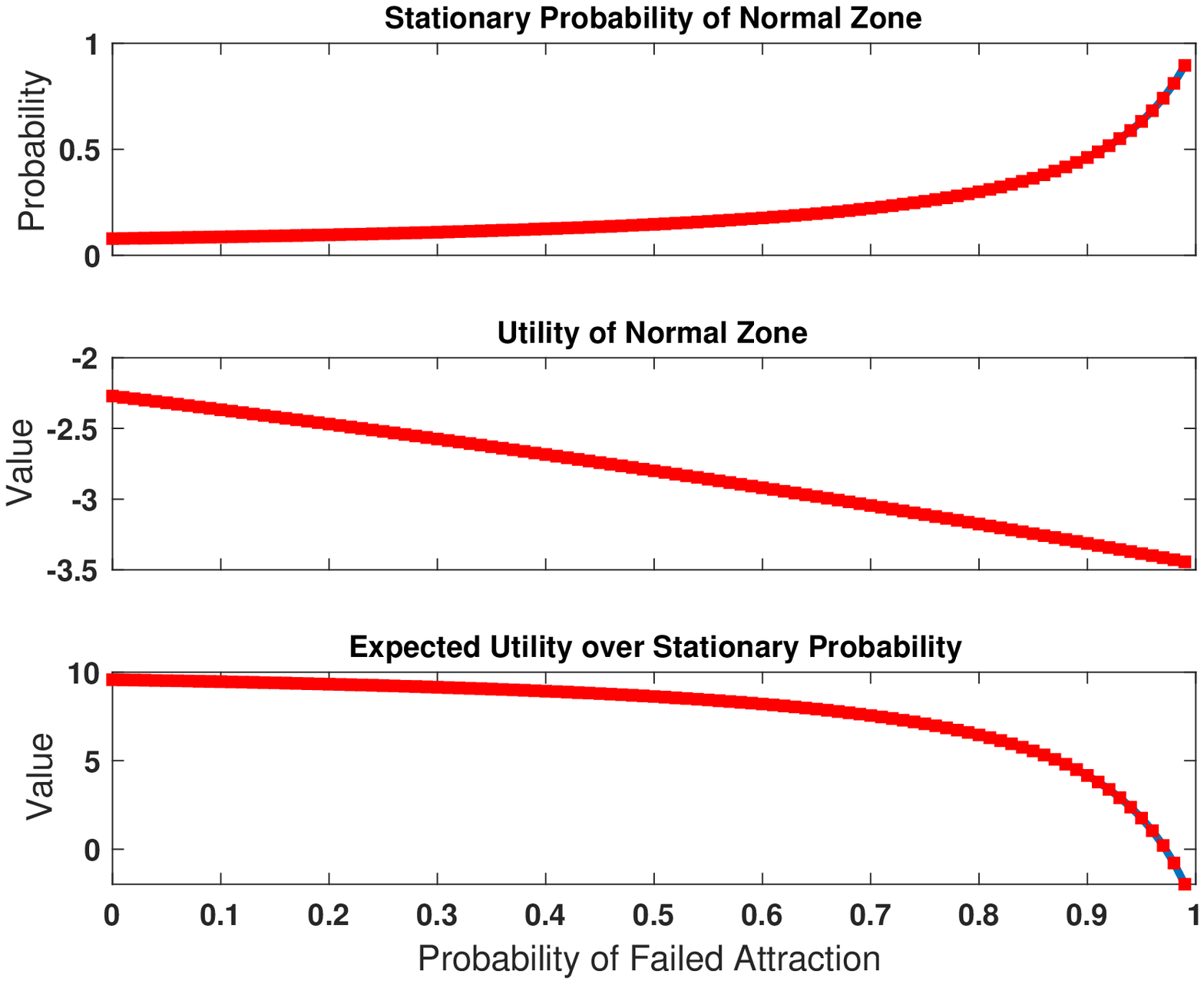}
\caption{
Three engagement criteria under different intelligence levels $p\in [0,1]$.
 \label{fig: advance}}
\endminipage
\end{figure}

In this section, we quantify three engagement criteria on attackers of different levels of persistence and intelligence in Fig. \ref{fig: persistency} and Fig. \ref{fig: advance}, respectively. 
The criteria are the stationary probability of normal zone $p_{i,s_{12}}(\infty), \forall i\in\mathcal{S}\setminus \{s_{13}\}$, the utility of normal zone $v(s_{12})$, and the expected utility over the stationary probability, i.e.,  $\sum_{j\in \mathcal{S}} p_{ij}(\infty)v(j), \allowbreak
\forall i\in\mathcal{S}\setminus \{s_{13}\}$. 

As shown in Fig. \ref{fig: persistency}, when the attacker is at the normal zone $i=s_{12}$ and the defender chooses action $a=a_A$, a larger $\lambda:=\lambda_{ij}(a_A), \forall j\in \{s_1,s_2,s_8\}$, of the exponential sojourn distribution indicates that the attacker is more inclined to respond to the honeypot attraction and thus less time is required to attract the attacker away from the normal zone. 
As the persistence level $\lambda$ increases from $0.1$ to $2.5$, the stationary probability of the normal zone decreases and the expected utility over the stationary probability increases, both converge to their stable values. The change rate is higher during $\lambda\in (0,0.5]$ and much lower afterward.  On the other hand, the utility loss at the normal zone decreases approximately linearly during the entire period $\lambda\in (0,2.5]$. 

As shown in Fig. \ref{fig: advance}, when the attacker becomes more advanced with a larger failure probability of attraction, i.e.,  $p:=p(j|s_{12},a_A), \forall j\in \{s_{12},s_{13}\}$, he can stay in the normal zone with a larger probability.  A significant increase happens after $p\geq 0.5$.  
On the other hand, as $p$ increases from $0$ to $1$, the utility of the normal zone reduces linearly, and the expected utility over the stationary probability remains approximately unchanged until $p\geq 0.9$. 

Fig. \ref{fig: persistency} and Fig. \ref{fig: advance} demonstrate that the expected utility over the stationary probability receives a large decrease only at the extreme cases of a high transition frequency and a large penetration probability. 
Similarly, the stationary probability of the normal zone remains small for most cases except for the above extreme cases. 
Thus, our policy provides a robust expected utility as well as a low-risk engagement over a large range of changes in the attacker's persistence and intelligence. 

\section{Reinforcement Learning of SMDP} 
\label{sec:reinforcementlearning}
Due to the absent knowledge of an exact SMDP model, i.e., the investigation reward, the attacker's transition probability (and even the network topology), and the sojourn distribution, the defender has to learn the optimal engagement policy based on the actual experience of the honeynet interactions. 
As one of the classical model-free reinforcement learning methods, the $Q$-learning algorithm for SMDP has been stated in \cite{bradtke1995reinforcement}, i.e., 
\begin{equation}
\label{eq:Qlearning}
\begin{split}
Q^{k+1}(s^k,a^k):=&(1-\alpha^k(s^k,a^k))Q^{k}(s^k,a^k)+ \alpha^k(s^k,a^k)[ \bar{r}_1(s^k,a^k,\bar{s}^{k+1})
\\
&
+\bar{r}_2(s^k,a^k)\frac{(1-e^{-\gamma \bar{\tau}^k})}{\gamma}-e^{-\gamma \bar{\tau}^k}\max_{a'\in \mathcal{A}(\bar{s}^{k+1})} Q^k(\bar{s}^{k+1},a')], 
\end{split}
\end{equation}
where $s^k$ is the current state sample,  $a^k$ is the current selected action, $\alpha^k(s^k,a^k)\in (0,1)$ is the learning rate, $\bar{s}^{k+1}$ is the observed state at next stage, $\bar{r}_1,\bar{r}_2$ is the observed investigation rewards, and $\bar{\tau}^k$ is the observed sojourn time at state $s^k$. 
When the learning rate satisfies $\sum_{k=0}^\infty \alpha^k(s^k,a^k)=\infty, \sum_{k=0}^\infty (\alpha^k(s^k,a^k))^2<\infty, \forall s^k\in \mathcal{S}, \forall a^k\in \mathcal{A}(s^k)$, and all state-action pairs are explored infinitely, $\max_{a'\in \mathcal{A}(s^k)} \allowbreak
Q^k(s^k,a'), k\rightarrow \infty$,  in \eqref{eq:Qlearning} converges to value $v(s^k)$ with probability $1$. 

At each decision epoch $k\in \{0,1,\cdots\}$, the action $a^k$ is chosen according to the $\epsilon$-greedy policy, i.e., the defender chooses the optimal action $arg\max_{a'\in \mathcal{A}(s^k)} Q^k(s^k,a')$ with a probability $1-\epsilon$, and a random action with a probability $\epsilon$. 
Note that the exploration rate $\epsilon\in (0,1]$ should not be too small to guarantee sufficient samples of all state-action pairs. The $Q$-learning algorithm under a pure exploration policy $\epsilon=1$ still converges yet at a slower rate. 

\begin{figure}
\RawFloats
\centering
\minipage[t]{0.80\textwidth}
\includegraphics[width=\textwidth]{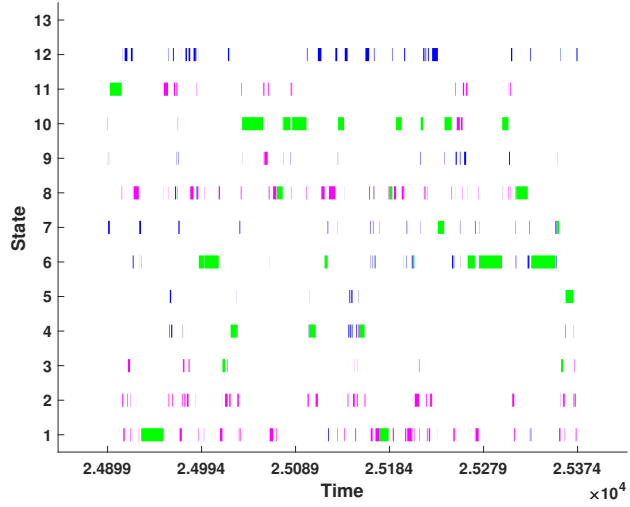}
\caption{
One instance of $Q$-learning on SMDP where the $x$-axis shows the sojourn time and the $y$-axis represents the state transition. 
The chosen actions $a_E,a_P,a_L,a_H$ are denoted in red, blue, purple, and green, respectively. 
 \label{fig: SamplePath}}
\endminipage
\end{figure}

In our scenario, the defender knows the reward of ejection action $a_A$ and $v(s_{13})=0$, thus does not need to explore action $a_A$ to learn it. 
We plot one learning trajectory of the state transition and sojourn time under the  $\epsilon$-greedy exploration policy in Fig. \ref{fig: SamplePath}, where the chosen actions $a_E,a_P,a_L,a_H$ are denoted in red, blue, purple, and green, respectively. 
If the ejection reward is unknown, the defender should be restrictive in exploring $a_A$ which terminates the learning process. Otherwise, the defender may need to  engage with a group of attackers who share similar behaviors to obtain sufficient samples to learn the optimal engagement policy. 

\begin{figure}
\RawFloats
\centering
\minipage[t]{0.48\textwidth}
\includegraphics[width=\textwidth]{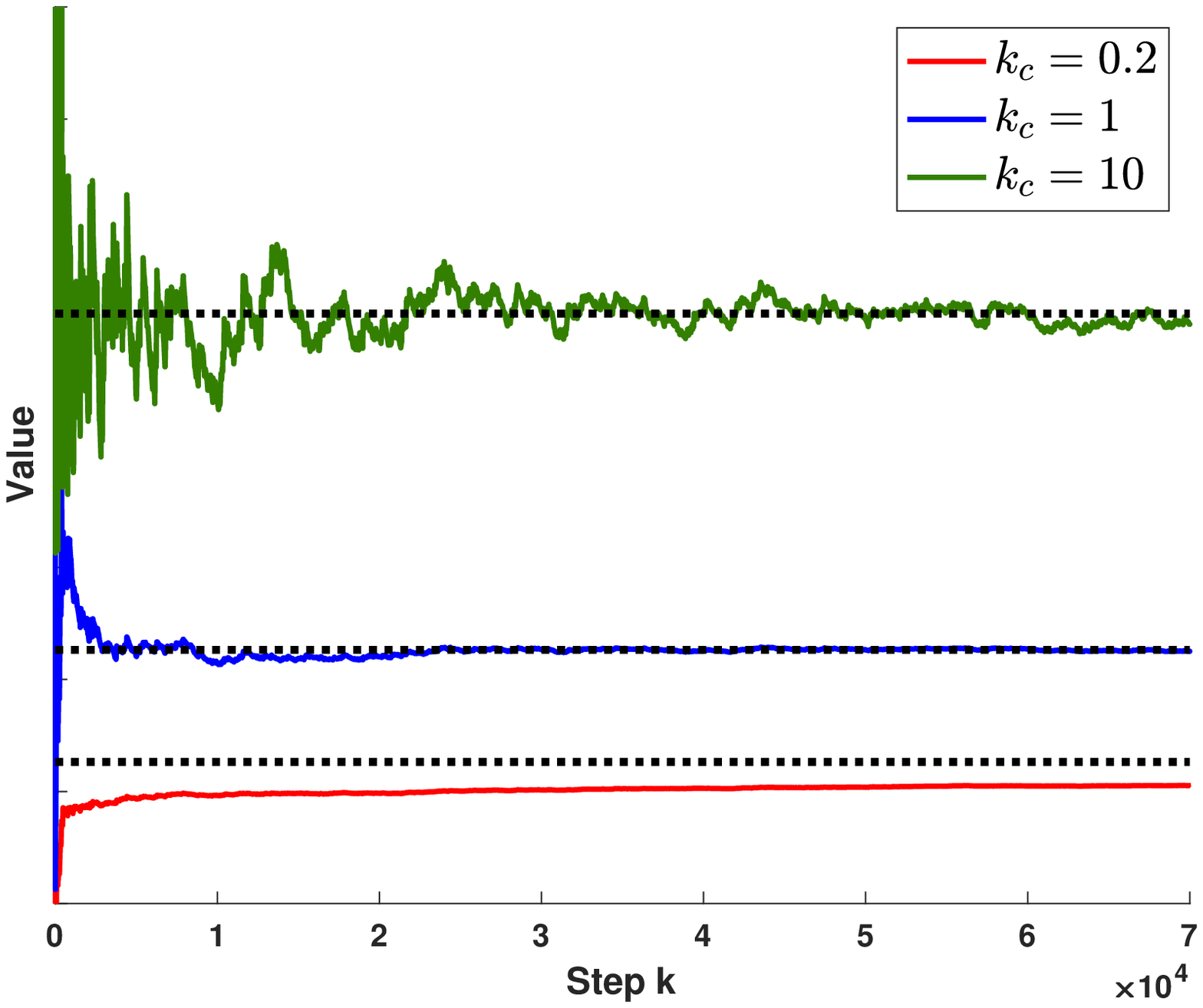}
\caption{
The convergence rate under different values of $k_c$. 
 \label{fig: compare_kc}}
\endminipage
\hfill
\minipage[t]{0.48\textwidth}
\includegraphics[width=\textwidth]{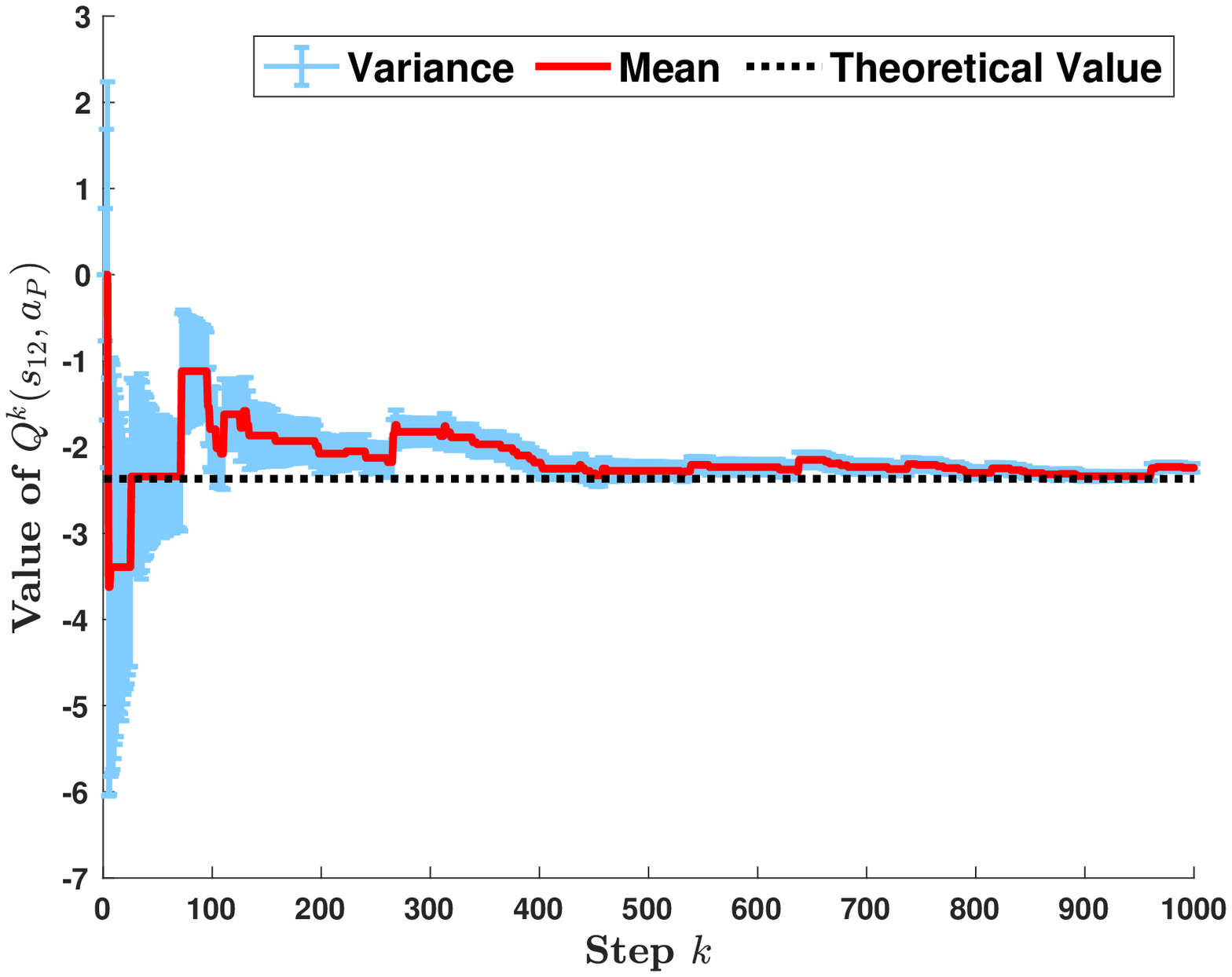}
\caption{
The evolution of the mean and the variance of $Q^k(s_{12},a_P)$. 
 \label{fig: Sample100}}
\endminipage
\end{figure}

In particular, we choose $\alpha^k(s^k,a^k)=\frac{k_c}{k_{\{s^k,a^k\}}-1+k_c}, \forall s^k\in \mathcal{S}, \forall a^k\in \mathcal{A}(s^k)$, to guarantee the asymptotic convergence, where $k_c \in (0,\infty)$ is a constant parameter and  $k_{\{s^k,a^k\}}\in \{0,1,\cdots\}$ is the number of visits to state-action pair $\{s^k,a^k\}$ up to stage $k$. 
We need to choose a proper value of $k_c$ to guarantee a good numerical performance of convergence in finite steps as shown in Fig. \ref{fig: compare_kc}. We shift the green and blue lines vertically to avoid the overlap with the red line and 
represent the corresponding theoretical values in dotted black lines. 
If $k_c$ is too small as shown in the red line, the learning rate decreases so fast  that new observed samples hardly update the $Q$-value and the defender may need a long time to learn the right value. 
However, if $k_c$ is too large as shown in the green line, the learning rate decreases so slow that new samples contribute significantly to the current $Q$-value. It causes a large variation and a slower convergence rate of $\max_{a'\in \mathcal{A}(s_{12})} Q^k(s_{12},a')$. 

We show the convergence of the policy and value under $k_c=1,\epsilon=0.2$, in the video demo (See URL: \url{https://bit.ly/2QUz3Ok}). 
In the video, the color of each node $n^k$ distinguishes the defender's action $a^k$ at state $s^k$ and the size of the node is proportional to $\max_{a'\in \mathcal{A}(s^k)} Q^k(s^k,a')$ at stage $k$. 
To show the convergence, we decrease the value of $\epsilon$ gradually to $0$ after $5000$ steps. 
 
Since the convergence trajectory is stochastic, we run the simulation for $100$ times and plot the mean and the variance of $Q^k(s_{12}, a_P)$ of state $s_{12}$ under the optimal policy $\pi(s_{12})=a_P$ in Fig. \ref{fig: Sample100}. The mean in red converges to the theoretical value in about $400$ steps and the variance in blue reduces dramatically as step $k$ increases. 

\subsection{Discussion}
\label{sec:Discussion}
In this section, we discuss the challenges and related future directions about reinforcement learning in the honeypot engagement. 
\subsubsection{Non-cooperative and Adversarial Learning Environment}
The major challenge of learning under the security scenario is that the defender lacks full control of the learning environment, which limits the scope of feasible reinforcement learning algorithms. 
In the classical reinforcement learning task, the learner can choose to start at any state at any time, and repeatedly simulate the path from the target state. 
In the adaptive honeypot engagement problem, however, the defender can remove attackers but cannot arbitrarily draw them to the target honeypot and force them to show their attacking behaviors because the true threat information is revealed only when attackers are unaware of the honeypot engagements.  
The future work could generalize the current framework to an adversarial learning environment where a savvy attacker can detect the honeypot and adopt deceptive behaviors to interrupt the learning process. 

\subsubsection{Risk Reduction during the Learning Period} 
Since the learning process is based on samples from real interactions, the defender needs to concern the system safety and security during the learning period. 
For example, if the visit and sojourn in the normal zone bring a significant amount of losses, we can use the SARSA algorithm to conduct a more conservative learning process than $Q$-learning.  
Other safe reinforcement learning methods are stated in the survey \cite{garcia2015comprehensive}, which are left as future work. 

\subsubsection{Asymptotic versus Finite-Step Convergence}
Since an attacker can terminate the interaction on his own, the engagement time with attacker may be limited. 
Thus, comparing to an asymptotic convergence of policy learning, the defender aims more to conduct speedy learning of the attacker's behaviors in finite steps, and meanwhile, achieve a good engagement performance in these finite steps. 

Previous works have studied the convergence rate  \cite{even2003learning} and the non-asymptotic convergence \cite{kearns1999finite,kearns2002near}  in the MDP setting.  For example,  \cite{even2003learning} have shown a relationship between the convergence rate and the learning rate of $Q$-learning, \cite{kearns1999finite} has provided the performance bound of the finite-sample convergence rate, and  \cite{kearns2002near} has proposed $E^3$ algorithm which achieves near-optimal with a large probability in polynomial time.  
However, in the honeypot engagement problem, the defender does not know the remaining steps that she can interact with the attacker because the attacker can terminate on his own. Thus, we cannot directly apply the $E^3$ algorithm which depends on the horizon time.  
Moreover, since attackers may change their behaviors during the long learning period, the learning algorithm needs to adapt to the changes of SMDP model quickly. 

In this preliminary work, we use the $\epsilon$-greedy policy for the trade-off of the exploitation and exploration during the finite learning time. The $\epsilon$ can be set at a relatively large value without the gradual decrease so that the learning algorithm persistently adapts to the changes in the environment. 
On the other hand, the defender can keep a larger discounted factor $\gamma$ to focus on the immediate investigation reward. 
If the defender expects a short interaction time, i.e.,  the attacker is likely to terminate in the near future, she can increase the discounted factor in the learning process to adapt to her  expectations. 
\subsubsection{Transfer Learning}
In general, the learning algorithm on SMDP converges slower than the one on MDP because the sojourn distribution introduces extra randomness. 
Thus, instead of learning from scratch, the defender can attempt to reuse the past experience with attackers of similar behaviors to expedite the learning process, which motivates the investigation of transfer learning in reinforcement learning \cite{taylor2009transfer}. Some side-channel information may also contribute to the transfer learning.


%

\section{Conclusion}
A honeynet is a promising active defense scheme.  
Comparing to traditional passive defense techniques such as the firewall and intrusion detection systems, the engagement with attackers can reveal a large range of Indicators of Compromise (IoC) at a lower rate of false alarms and missed detection. 
However, the active interaction also introduces the risks of attackers identifying the honeypot setting, penetrating the production system, and a high implementation cost of persistent synthetic traffic generations. 
Since the reward depends on honeypots' type, the defender aims to lure the attacker into the target honeypot in the shortest time. 
To satisfy the above requirements of security, cost, and timeliness, we leverage the Semi-Markov Decision Process (SMDP) to model the transition probability, sojourn distribution, and investigation reward. 
After transforming the continuous time process into the equivalent discrete decision model, we have obtained long-term optimal policies that are risk-averse, cost-effective, and time-efficient. 

We have theoretically analyzed the security metrics of the \textit{occupancy distribution}, \textit{attraction efficiency}, and \textit{average engagement efficiency} based on the transition probability and the probability density function of the first passenger time. 
The numerical results have shown that the honeypot engagement can engage the attacker in the target honeypot with a large probability and in a desired speed. 
In the meantime, the penetration probability is kept under a bearable level for most of the time. 
The results also demonstrate that it is a worthy compromise of the immediate security to allow a small penetration probability so that a high investigation reward can be obtained in the long run.  

Finally, we have applied reinforcement learning methods on the SMDP  in case the defender can not obtain the exact model of the attacker's behaviors.  
Based on a prudent choice of the learning rate and exploration-exploitation policy, 
 we have achieved a quick convergence rate of the optimal policy and the value. Moreover, the variance of the learning process has decreased dramatically with the number of observed samples. 

%
%
%
 \bibliographystyle{splncs04}
\bibliography{APT}
\end{document}